\documentclass[9pt,shortpaper,twoside,web]{ieeecolor}
\usepackage{generic}
\usepackage{cite}
\usepackage{amsmath,amssymb,amsfonts}
\usepackage{algorithmic}
\usepackage{graphicx}
\usepackage{textcomp}
\usepackage{booktabs}  
\usepackage{graphicx}  
\usepackage{array}   
\usepackage{float}
\usepackage{threeparttable} 
\def\BibTeX{{\rm B\kern-.05em{\sc i\kern-.025em b}\kern-.08em
    T\kern-.1667em\lower.7ex\hbox{E}\kern-.125emX}}

\begin{document}
\title{DLGE: Dual Local-Global Encoding for Generalizable Cross-BCI-Paradigm}
\author{Jingyuan Wang, \IEEEmembership{Graduate Student Member, IEEE}, Junhua Li, \IEEEmembership{Senior Member, IEEE}
\thanks{Jingyuan Wang and Junhua Li are with the School of Computer Science and Electronic Engineering, University of Essex, CO4 3SQ Colchester, U.K.}
\thanks{Corresponding author: Junhua Li (e-mail: junhua.li@essex.ac.uk).}}

\maketitle

\begin{abstract}
Deep learning models have been frequently used to decode a single brain-computer interface (BCI) paradigm based on electroencephalography (EEG). It is challenging to decode multiple BCI paradigms using one model due to diverse barriers, such as different channel configurations and disparate task-related representations. In this study, we propose Dual Local-Global Encoder (DLGE), enabling the classification across different BCI paradigms. To address the heterogeneity in EEG channel configurations across paradigms, we employ an anatomically inspired brain-region partitioning and padding strategy to standardize EEG channel configuration. In the proposed model, the local encoder is designed to learn shared features across BCI paradigms within each brain region based on time-frequency information, which integrates temporal attention on individual channels with spatial attention among channels for each brain region. These shared features are subsequently aggregated in the global encoder to form respective paradigm-specific feature representations. Three BCI paradigms (motor imagery, resting state, and driving fatigue) were used to evaluate the proposed model. The results demonstrate that our model is capable of processing diverse BCI paradigms without retraining and retuning, achieving average macro precision, recall, and F1-score of 60.16\%, 59.88\%, and 59.56\%, respectively. We made an initial attempt to develop a general model for cross-BCI-paradigm classification, avoiding retraining or redevelopment for each paradigm. This study paves the way for the development of an effective but simple model for cross-BCI-paradigm decoding, which might benefit the design of portable devices for universal BCI decoding. 
\end{abstract}


\section{Introduction}
\label{sec:introduction}
\IEEEPARstart{B}RAIN-COMPUTER interface (BCI) has emerged in recent decades as a promising communication system, which enables a direct pathway between the human brain and external devices or environments \cite{gong_deep_2022}. Among various neurophysiological signals adopted in BCI, electroencephalography (EEG) is popular with its excellent temporal resolution, low cost, ease of use, and non-invasive advantage \cite{varbu_past_2022,flesher_brain-computer_2021, ehrlich_closed-loop_2019,sun_adapteeg_2025}. By the use of EEG, different BCI paradigms have been developed for diverse applications, such as motor imagery decoding \cite{liu_fine-grained_2025,yu_deep_2024,autthasan_min2net_2022,hsu_eeg-channel-temporal-spectral-attention_2023,li_design_2013}, resting-state monitoring \cite{torkamani-azar_prediction_2020}, and fatigue detection \cite{guo_driver_2023,shahbakhti_simultaneous_2022,min_driver_2017,wang_novel_2018,harvy_reliability_2022}. Currently, decoding models are usually designed based on a particular BCI paradigm. Even if the same model can be used, it must be retrained when applied to a new BCI paradigm. This is mainly due to two crucial challenges that hinder cross-BCI-paradigm classification. These two challenges are the lack of consistent configurations of EEG channels across BCI paradigms and diverse cognitive or behavioural tasks that result in varying brain activation patterns and EEG characteristics. Both challenges collectively make it difficult to design a structured and generalizable model capable of effective cross-BCI-paradigm classification.

A large body of EEG decoding models relies on convolutional neural networks (CNNs), which are well suited for capturing local spatial and temporal features. Pioneering works such as ConvNet \cite{schirrmeister_deep_2017}, EEGNet \cite{lawhern_eegnet_2018}, and TSception \cite{ding_tsception_2023} demonstrate high performance on a single paradigm. These models operate under the assumption of a fixed number of EEG channels arranged in a predefined order and cannot be directly used for the case of cross-BCI-paradigm, in which channel configurations differ from each other. Extending these models by incorporating attention mechanisms for global modeling, Transformer-based models, including Conformer \cite{song_eeg_2023} and Deformer \cite{ding_eeg-deformer_2025} were proposed. While these models achieve excellent performance on a single paradigm, they still employ CNN-based encoders or positional embeddings tied to specific channel configurations, implying that these models require a prior knowledge of channel configurations, making them unable to generalize to new paradigms without retraining. Moreover, to address these limitations in a more flexible manner, graph-based models such as EmT \cite{ding_emt_2025} and HN-DGST \cite{cheng_hybrid_2024} leverage graph convolutional networks (GCNs) to represent inter-channel relationships. However, these models require a fixed adjacency matrix based on the original channel configuration, which again assumes consistent channel configuration across paradigms. Consequently, these models exhibit limited adaptability to the heterogeneous channel configuration that characterize cross-BCI-paradigm.

Besides the challenge of channel configuration heterogeneity, another major limitation of current EEG models is their inability to generalize across BCI paradigms involving different cognitive or behavioural tasks. Although this limitation is acknowledged as a significant challenge, most studies focus on the relatively easier tasks of improving generalization across subjects or sessions within the same BCI paradigm, where the task types and channel configurations remain consistent. For example, several studies aimed to enhance cross-subject generalization in a single paradigm \cite{sun_adapteeg_2025}. In an earlier work, Li et al. explored a wide range of linear and nonlinear EEG features to improve cross-subject decoding \cite{li_exploring_2018}. Shen et al. proposed a contrastive learning method CLISA for subject-invariant representation learning using convolutional alignment \cite{shen_contrastive_2023}. Liu et al. employed a domain-adaptive multi-branch capsule network DA-CapsNet to improve subject transferability \cite{liu_da-capsnet_2024}. Similarly, an ensemble of neural networks optimized by evolutionary programming (EPNNE) was introduced to boost cross-subject classification performance \cite{zhang_evolutionary_2024}. Other studies have tackled cross-session variability, which is caused by intra-subject changes over time due to factors such as tiredness, electrode shift, or ambient change. For instance, the JCSFE \cite{peng_cross-session_2023} utilized a semi-supervised framework to jointly learn session-invariant and session-specific features. To jointly handle cross-subject and cross-session variability, a multi-source domain adaptation network was proposed, which aligns both shared and specific representations across domains \cite{she_multisource_2023}. More recently, FMLAN \cite{yu_fmlan_2025} introduced a fine-grained mutual learning adaptive network that simultaneously enhances emotion-discriminative feature learning and domain alignment across both subjects and sessions. While these methods have advanced subject/session-level generalization, they are limited by their assumption of a consistent task type. In practical applications, BCI paradigms differ widely in diverse cognitive or behavioral tasks, such as motor imagery, resting state, or fatigue state, each engaging distinct neural activity \cite{zich_real-time_2015, wang_phase_2006,li_self-regulation_2023}. For example, motor imagery paradigm typically involves sensorimotor cortex and strengthens neural activities in mu and beta frequency bands \cite{yu_effects_2022}. In contrast, the driving fatigue paradigm is associated with frontal pole and frontal regions in the lower alpha band (8–10 Hz) \cite{sun_functional_2014,wang_driving_2021}. Resting state paradigm exhibits a distinct right hemisphere topography in the theta band \cite{barry_eeg_2007}.

The aforementioned challenges are barriers to the development of cross-BCI-paradigm classification. The existing models are not able to achieve reasonable performance unless they are retrained or redeveloped to adapt to a new paradigm. In order to develop a generalizable classification model fitting different models, the model must be compatible with diverse channel configurations and different tasks across BCI paradigms. To address this issue, it is crucial to understand the major challenges in cross-BCI-paradigm classification, which can be summarized as follows:

\begin{itemize}
\item
Lack of effective solutions for handling channel heterogeneity cross-BCI-paradigm. Previously proposed models are inherently restricted by architectural constraints, preventing them from processing EEG signals with varying configuration of channels. Even for models that are structurally capable of accepting heterogeneous inputs, channel inconsistency often introduces noise and impairs the model’s ability to learn reliable and discriminative representations. Therefore, it is crucial to develop a module that can address inconsistent EEG channel configurations across different paradigms while reducing noise induced by channel heterogeneity.
\item
Lack of appropriate strategies to address interference between various types of tasks. In scenarios involving multiple types of tasks, the neural activity associated with different task types often occurs in distinct frequency bands. This often results in a model trained to perform well on one task type may fail to generalize effectively to others. Therefore, it is essential to develop mechanisms that alleviate this interference, enabling the model to learn more stable and task-related representations across different types of EEG tasks.
\end{itemize}

To tackle the challenge posed by inconsistent channel configurations across paradigms, we employ an anatomically inspired brain-region partitioning and padding strategy. By standardizing EEG channels into a predefined anatomical brain-region configuration and padding missing channels accordingly, we transformed different channel configurations of each paradigm into a unified form. This step enables the model to process heterogeneous channel configurations from different paradigms in a unified form. In addition, motivated by the success of spatio-spectral modeling across brain regions in EEG-based classification \cite{goh_spatiospectral_2018, wang_multikernel_2022}, our proposed model incorporates a dual-level encoder framework to learn more stable and task-related representations. The Local Brain Encoder (LBE) extracts core regional features that are consistent across paradigms.
Subsequently, the Global Brain Encoder (GBE) aggregates these shared fundamental regional features into high-level representations that are specific to each task. This dual decoding framework allows the model to learn both shared and task-dependent characteristics across diverse paradigms. The main contributions of this work are summarized as follows:

\begin{itemize}
\item
We propose an anatomical inspired
brain-region partitioning and padding strategy to address inconsistencies in channel configurations across paradigms. Our approach allows the model to handle varied channel configurations across paradigms in a consistent manner, without any dependence on paradigm-specific adjustments or retraining.
\item
We design a dual-level encoder to mitigate task-specific interference and capture both shared and task-specific patterns. The LBE learns fundamental features within each brain region using spatial and temporal attention, while the GBE aggregates these features into high-level task-specific representations across paradigms.
\item
We demonstrate the effectiveness of the proposed model DLGE is effective on the classification across BCI paradigms according to the evaluation results based on three different BCI paradigms.
\end{itemize}

The remainder of this paper is organized as follows. Section II introduces the BCI paradigms used in this study and presents a comprehensive description of the proposed DLGE model. Section III reports the experimental results. Finally, Section IV concludes the paper and outlines directions for future research.

\section{Methods}
\subsection{Datasets}
To evaluate the feasibility of our model in cross-BCI-paradigm classification, three paradigms are included, which are the motor imagery paradigm \cite{tangermann_review_2012}, the resting state paradigm \cite{torkamani-azar_prediction_2020}, and the driving fatigue paradigm \cite{min_driver_2017}. A setting comparison of the three BCI paradigms is summarized in Table~\ref{tab:datasets}. 

\begin{table*}[t]
\centering
\caption{Setting Comparison of Three BCI Paradigms}
\label{tab:datasets}
\begin{tabular}{lccccc}
\toprule
\textbf{Paradigm} & \textbf{\#Subjects} & \textbf{\#Channels} & \textbf{Sampling Rate (Hz)} & \textbf{\#Classes} & \textbf{Class Labels}   \\
\midrule
Motor Imagery Paradigm        & 9                   & 22                   & 250          & 4      & Left Hand, Right Hand, Feet, Tongue            \\
Resting State Paradigm        & 10                  & 64                   & 2048           & 2     & Eyes Open, Eyes Closed             \\
Driving Fatigue Paradigm     & 12                  & 32                   & 1000        & 2          & Alertness, Fatigue           \\
\bottomrule
\end{tabular}
\end{table*}

\subsubsection{Motor Imagery Paradigm}

The data of the motor imagery paradigm are from BCI Competition IV-2a dataset \cite{tangermann_review_2012}. There are four tasks of motor imagery (i.e., left hand, right hand, both feet, and tongue). Nine subjects participated in the experiment, each of whom performed two sessions on separate days. Each participant completed two sessions. Each session consisted of six runs, with short breaks between them. Each run included 48 trials (4 tasks $\times$ 12 trials), resulting in 288 trials per session. EEG signals were recorded using 22 active Ag/AgCl electrodes positioned according to the international 10–20 system, with a sampling rate of 250 Hz.
\subsubsection{Resting State Paradigm}
The resting-state paradigm utilizes the Resting-State Dataset \cite{torkamani-azar_prediction_2020}, which includes under eyes open and eyes closed conditions. EEG signals were recorded using a 64-channel Biosemi ActiveTwo system, with electrodes placed according to the international 10–10 system and a sampling rate of 2048 Hz. Each subject completed two sessions corresponding to the open and closed eyes conditions. Each session lasted 2.5 minutes. Ten healthy individuals participated in the paradigm, including six females and four males, with a mean age of 30.25 ± 6.95 years and an age range from 22 to 45.5 years. In addition, all subjects were right-handed, had normal or corrected normal vision, and had not taken any hypnotic medication during the experiment.

\subsubsection{Driving Fatigue Paradigm}
For the driving fatigue paradigm, we employ the Driver Fatigue Dataset \cite{min_driver_2017}, which contains EEG signals collected under two mental states: alertness and fatigue. To ensure data quality, all 12 healthy male participants (aged 12 to 24 years) were instructed to maintain adequate sleep and refrain from taking stimulants or medications before the experiment. EEG signals were recorded using a 40-channel Neuroscan amplifier, with 32 active Ag/AgCl electrodes placed according to the international 10–20 system. For each subject, 5-minute EEG segments were collected at a sampling rate of 1000 Hz under both alertness and fatigue conditions.

\begin{figure*}[t]
\centerline{\includegraphics[width=0.9\textwidth]{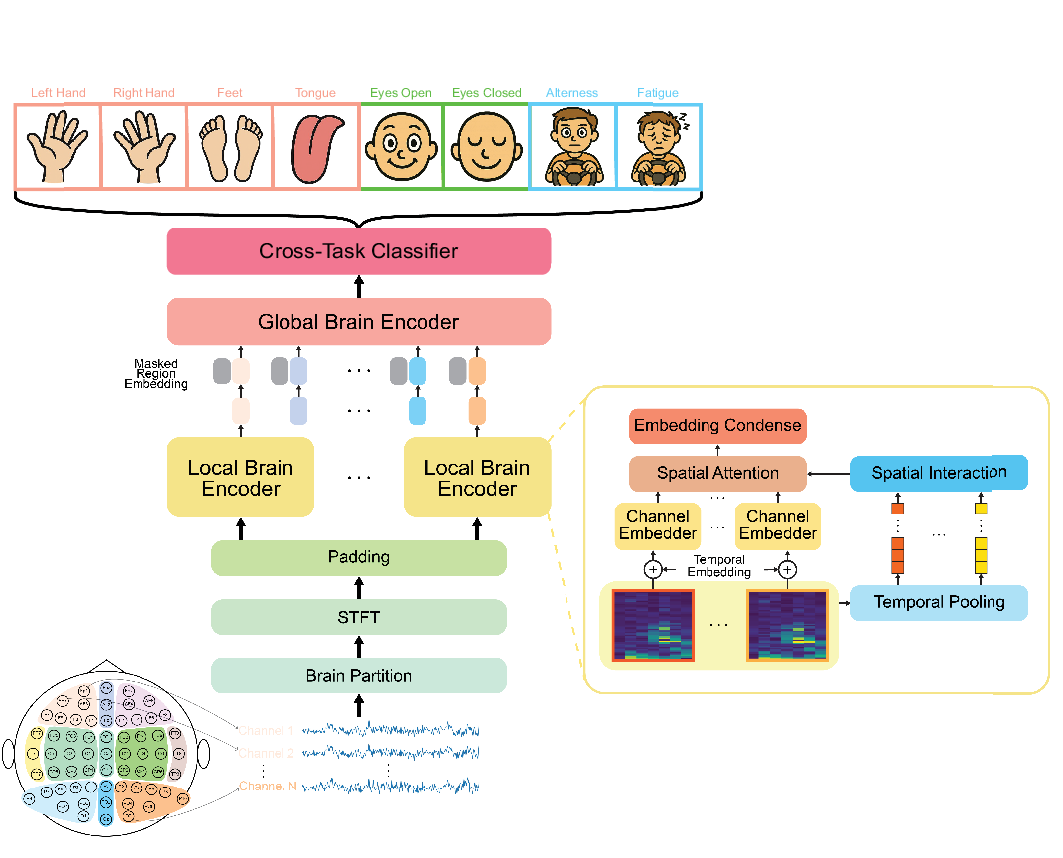}}
\caption{The schematic of the proposed Dual Local-Global Encoder (DLGE) model for cross-BCI-paradigm classification. Each padded brain region is used as the input to the local brain encoder (LBE) that will extract fundamental shared feature within each brain by channel embedder and spatial attention. The input of channel embedder is the time–frequency features of each channel. As for the spatial attention, the input is spatial-frequency features of each brain region. Then the global brain encoder (GBE) is applied to integrate fundamental regional features into task-specific high level representaions. After that, the cross-task classifier is employed to obtain classification results.}
\label{fig:model}
\end{figure*}

\subsection{Data Standardization}
 All EEG data from the three paradigms are resampled to 250 Hz and band-pass filtered between 0.5–45 Hz. Each trial is segmented into non-overlapping 3-second segments, which serve as individual samples in this study. In the motor imagery paradigm, each subject complete 576 trials (144 per class), yielding a total of 5,184 samples (1296 per class) from 9 subjects across four classes. The resting-state paradigm involve 5 minutes per subject. After segmentation, around 1000 samples (500 per class) are obtained in total. For the driving fatigue paradigm, segmentation of 5-minute sessions under alertness and fatigue states resulted in 1,200 samples per state, totaling 2,400 samples across subjects. Finally, we shuffle all the samples.

\subsection{Dual Local-Global Encoder}
As shown in Fig.~\ref{fig:model}, our proposed model consists of four main components. The first component addresses the heterogeneity of channel configuration, caused by different settings across BCI paradigms, by introducing an anatomically inspired brain region partitioning and padding strategy. The second component is the LBE, which is designed to capture fundamental regional features shared across paradigms. Both local spatial attention and temporal attention are embedded into the LBE. The third component, named GBE, is designed to integrate fundamental regional features into high-level feature representations associated with each task. The final component is the cross-task classifier, which performs the classification based on the inputted high-level feature representations.

Let $\mathbf{X} \in \mathbb{R}^{C \times T}$ denotes a sample with $C$ channels and $T$ time points. EEG channels are grouped into 11 sets (corresponding to 11 brain regions, $R = 11$) according to their anatomically spatial locations. These regions are the left frontal region, middle frontal region, right frontal region, left temporal region, left central region, middle motor cortex region, right motor cortex region, right temporal region, left parietal-occipital region, middle parietal-occipital region, and right parietal-occipital region. The regional grouping can be expressed as follows
\begin{equation}
    \mathbf{X}=\left\{\mathbf{X}^{(1)}, \mathbf{X}^{(2)}, \ldots, \mathbf{X}^{(R)}\right\}, \quad \mathbf{X}^{(r)} \in \mathbb{R}^{C_{r} \times T}.
\end{equation}

 We subsequently extract time-frequency features from each EEG channel using Short-Time Fourier Transform (STFT) with the sliding time window of 250 data points and an overlap rate of 50\%. After STFT, we obtain a feature tensor $\mathbf{\hat{X}}^{(r)} \in \mathbb{R}^{C_{r} \times F \times T'}$, where $F$ denotes the number of frequencies and $T'$ is the number of time points.

In order to unify the different numbers of channels across paradigms, we align the channels within a region $r$ to a standardized configuration of $\tilde{C}= 9$ channels. To align with the standardized channel configuration, any missing channels in a paradigm are padded with zeros.
\begin{equation}
    \forall r \in\{1, \ldots, R\}, \quad \mathbf{\tilde{{X}}}^{(r)} \in \mathbb{R}^{\tilde{C} \times F \times T'}.
\end{equation}

\begin{figure*}[t]
\centerline{\includegraphics[width=0.9\textwidth]{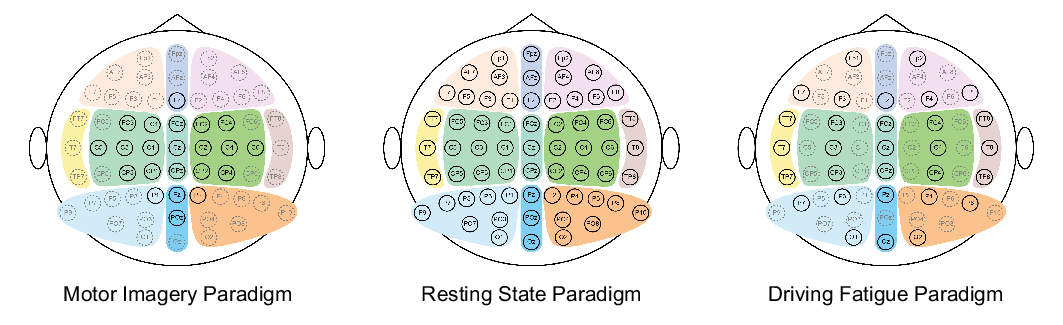}}
\caption{Illustration of the recorded electrode layouts and the padded electrodes for each paradigm, as well as the brain region partition in our study. Solid circles represent electrodes actually used for data recording, while dashed circles represent the missing electrodes that are padded in our study. Background colors indicate the partitioned brain regions.}

\label{fig:region_partition}
\end{figure*}

Therefore, each $\mathbf{\tilde{{X}}}^{(r)}$ contains the time-frequency representations of all valid channels within a specific brain region, with zero padding applied to align with the standardized channel configuration. The paradigms channel configurations, brain region partitioning strategy and the corresponding channel padding scheme are illustrated in Fig.~\ref{fig:region_partition}, which demonstrates the number of electrodes assigned to each region and the zero-padding strategy.

\subsubsection{Local Brain Encoder}
The Local Brain Encoder is designed to extract essential fundamental regional features that were shared across different paradigms within each brain region. It combines local spatial attention with temporal attention mechanisms to enhance its ability to capture shared fundamental local regional representations.
 
Initially, each channel is fed into the Channel Embedder, which is utilized to extract the representations of all valid (non-padded) EEG channels within each brain region. The Channel Embedder follows a conventional Transformer architecture, but employs single-head self-attention instead of multi-head attention in our implementation. 
This design preserves the natural frequency structure of EEG signals, as multi-head attention would split the frequency dimension across different heads, potentially breaking the meaningful relationships between frequency bands. Each channel within a region is treated as a sequence of tokens along the temporal axis. Specifically, the time-frequency matrix $\mathbf{X}^{(r)}_{i} \in \mathbb{R}^{F \times T'}$ of the $i$-th channel in $r$-th brain region is transposed and then incorporate temporal positional embeddings to encode temporal information as:

\begin{equation}
    \mathbf{Z}_{i}^{(r)}=\mathrm{PE}+\left(\mathbf{\tilde{X}}_{i}^{(r)}\right)^{\top} \in \mathbb{R}^{T^{\prime} \times F},
\end{equation}
where $\mathrm{PE} \in \mathbb{R}^{T' \times F}$ is a learnable temporal positional embedding. Channel Embedder processes each $\mathbf{Z}_{i}^{(r)}$ to yield the channel-level feature as
\begin{equation}
\mathbf{h}_{i}^{(r)}=\operatorname{Transformer}\left(\mathbf{Z}_{i}^{(r)}\right), \quad \bar{\mathbf{h}_{i}}^{(r)}=\frac{1}{T^{\prime}} \sum_{t=1}^{T^{\prime}} \mathbf{h}_{i}^{(r)}[t].
\end{equation}

Next, to capture inter-channel relationships within a brain region, we compute pairwise channel correlations. First, we apply temporal pooling on each region as
\begin{equation}
    \mathbf{v}^{(r)}=\operatorname{MeanPool}_{t}\left(\mathbf{\tilde{X}}^{(r)}\right) \in \mathbb{R}^{\tilde{C} \times F}.
\end{equation}
Then, following the attention mechanism, we project the pooled vectors into queries and keys as
\begin{equation}
    \mathbf{q}^{(r)}=\mathbf{W}^{q}\mathbf{v}^{(r)}, \quad \mathbf{q}^{(r)}=\mathbf{W}^{k}\mathbf{v}^{(r)},
\end{equation}
where $\mathbf{W}^{q}, \mathbf{W}^{k}\in \mathbb{R}^{F \times d_{\text{local}}}$. The attention scores across all channels in the region are computed as
\begin{equation}
    \mathbf{a}^{(r)}=\operatorname{softmax}\left(\frac{\mathbf{q}^{(r)} \mathbf{k}^{(r)\top}}{\sqrt{d_{\text{local}}}}\right) \in \mathbb{R}^{\tilde{C} \times \tilde{C}}.
\end{equation}
This procedure is referred to as Spatial Interaction, which quantifies the channel-wise relevance within a region. Then, the attention scores are used to reweight the channel embeddings as
\begin{equation}
    \tilde{\mathbf{h}}^{(r)}=\mathbf{a}^{(r)} \bar{\mathbf{h}}^{(r)}\in \mathbb{R}^{\tilde{C} \times d_{\text{local}}}.
\end{equation}
This step is referred to as spatial attention. The final representation for the region $r$ is obtained by applying a feed-forward network (FFN) followed by a channel-wise pooling operation to condense and aggregate the representations of all channels within the region as
\begin{equation}
    \mathbf{z}^{(r)}=\frac{1}{C^{\prime}} \sum_{i=1}^{C^{\prime}} \operatorname{FFN}\left(\tilde{\mathbf{h}}^{(r)}_{i}\right) \in \mathbb{R}^{d_{\text{local}}},
\end{equation}
where $C^{\prime}$ is the number of valid channels. This step is referred to as Embedding Condensation.

\subsubsection{Global Brain Encoder}
The Global Brain Encoder is designed to integrate fundamental regional features into comprehensive high-level representations tailored for each specific task. For each brain region, we firstly add a learnable masked region embedding to its representation, which encodes the positional information of the region. The resulting region-wise representations are then passed into the Global Brain Encoder, which is implemented as a standard multi-head attention Transformer. The Transformer follows the general formulation as
\begin{equation}
\mathbf{H}=\operatorname{Transformer}\left(\mathbf{z}\right)\in \mathbb{R}^{R \times d_{\text{global}}}.
\end{equation}
To prevent noise introduced by padded zeros in brain regions lacking valid channels during the self-attention computation, the attention scores corresponding to these padded positions are assigned a value of negative infinity before the softmax operation. This ensures that after applying softmax, the attention weights for the padded channels are effectively zeroed out, thereby eliminating their influence on the attention mechanism. The output is aggregated via mean pooling over the token (region) dimension to obtain a high-level representation $\bar{\mathbf{H}}$ as
\begin{equation}
\bar{\mathbf{H}}=\frac{1}{R^{\prime}} \sum_{r=1}^{R^{\prime}} \mathbf{H}[r],
\end{equation}
where $R^{\prime}$ is the number of valid regions.

\subsubsection{Cross-Task Classifier}
In the final step, the cross-task classifier consists of two fully-connected layers with ReLU activation in between. It takes the high-level representation as input and outputs the final prediction after $Softmax$ function as
\begin{equation}
    \hat{\mathbf{y}} = \mathrm{Softmax}\left(\mathrm{FC}_2\left( \mathrm{ReLU}\left( \mathrm{FC}_1\left( \bar{\mathbf{H}} \right) \right) \right)\right).
\end{equation}

\subsection{Evaluation}
For reliable performance evaluation, we addressed class imbalance by implementing a class-weighted cross-entropy loss function during training. This approach prevents the model from developing bias toward majority classes. Our experimental design utilized five-fold cross-validation, with the dataset partitioned into five folds maintaining approximately equal class proportion.

To provide a thorough evaluation of the model's performance, we adopted macro precision, macro recall, and macro F1-score obtained via five-fold cross-validation as evaluation metrics. The formulas are given as
\begin{equation}
    \text { Macro Precision }=\frac{1}{C_{n}}\sum_{c=1}^{C_{n}} \cdot \frac{\mathrm{TP}_{c}}{\mathrm{TP}_{c}+\mathrm{FP}_{c}},
\end{equation}
\begin{equation}
    \text { Macro Recall }=\frac{1}{C_{n}}\sum_{c=1}^{C_{n}} \cdot \frac{\mathrm{TP}_{c}}{\mathrm{TP}_{c}+\mathrm{FN}_{c}},
\end{equation}
\begin{equation}
    \text { Macro F1-Score }=\frac{1}{C_{n}}\sum_{c=1}^{C_{n}} \cdot \frac{2 \cdot \text {Precision}_{c} \cdot \text {Recall}_{c}}{\text {Precision}_{c}+\text {Recall}_{c}},
\end{equation}
where $C_{n}$ is the number of classes, $\mathrm{TP}_{c}, \mathrm{FP}_{c}, \mathrm{FN}_{c}$ denote the number of true positives, false positives, and false negatives for class $c$, respectively.
This macro averaging scheme ensures that the evaluation reflects the model’s ability to perform well across all classes, including those with fewer samples. 

\section{Results}
\subsection{Classification Performance}

To demonstrate the effectiveness of the DLGE, we evaluate our model on a new dataset that we constructed by shuffling all samples from three different paradigms. The macro precision, macro recall, and macro F1-score achieved in each fold are presented in Table~\ref{tab:results}.

\begin{table}[t]

\caption{The results of DLGE on cross-BCI-paradigm classification}
\label{tab:results}
\begin{center}
  \begin{tabular}{cccc}
    \toprule
    Fold & Mac. Prec. & Mac. Rec. & Mac. F1 \\
    \midrule
     1 & 58.58\% & 58.58\% & 58.39\% \\
     2 & 58.72\% & 58.89\% & 58.37\% \\
     3 & 60.12\% & 59.43\% & 59.39\% \\
     4 & 61.34\% & 61.00\% & 60.68\% \\
     5 & 62.06\% & 61.49\% & 60.97\% \\
    
    \textbf{Mean}& \textbf{60.16\%} & \textbf{59.88\%} & \textbf{59.56\%} \\
    \bottomrule
\end{tabular}
\end{center}
\vspace{1mm}
\scriptsize 
Mac. Prec. = Macro Precision; Mac. Rec. = Macro Recall; Mac. F1 = Macro F1-Score.

\end{table}

The evaluation results demonstrate the feasibility and effectiveness of the proposed  DLGE model for cross-BCI-paradigm classification. It achieved an average macro precision of 60.16\%, macro recall of 59.88\%, and macro F1-score of 59.56\% according to the five-fold cross-validation on the data of the three different BCI paradigms. In particular, the relatively small standard deviation between folds (approximately 1~2\%) reflects consistent performance across data splits, demonstrating the robustness of the proposed approach. These consistent results across folds suggest that DLGE is feasible to various paradigms, benefiting from its local-global high level task-specific modeling and brain-region-based EEG standardization. 

\subsection{Ablation Study}
\begin{figure}[!t] 
\centerline{\includegraphics[width=\columnwidth]{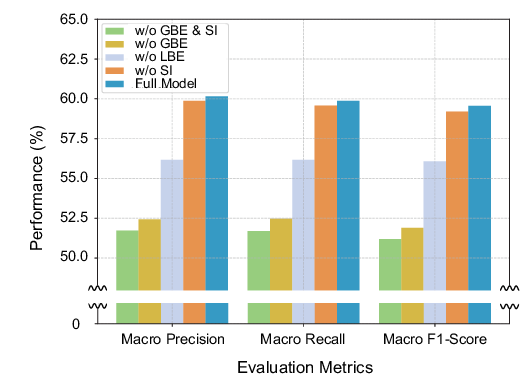}}
\caption{Ablation study results of the proposed Dual Local-Global Encoder (DLGE) model across three BCI paradigms. We evaluate the impact of removing key architectural components, which are global brain encoder (GBE), local brain encoder (LBE) and its internal spatial interaction (SI) module, SI module, and both GBE and SI module. The removal of GBE leads to the most significant performance drop across all evaluation metrics, underscoring its critical role in modeling global context. Excluding LBE+SI results in moderate degradation, highlighting the importance of localized feature extraction. The absence of only SI causes marginal impact, suggesting its auxiliary role in enhancing regional interactions. }
\label{fig:ab}
\end{figure}

To better understand the contribution of each module in DLGE to the overall performance, we conducted an ablation study. Specifically, we conducted ablation studies by comparing four model configurations including DLGE without the GBE, DLGE without both the LBE and its internal Spatial Interaction (SI) module, DLGE without the SI module, and DLGE without both GBE and SI module, as illustrated in Fig.~\ref{fig:ab}.

The Fig. 3 shows that the removal of the GBE has the most significant negative impact on our model performance. Specifically, the macro precision dropped by 7.73\%, macro recall by 7.4\%, and macro F1-Score by 7.67\%, highlighting the critical role of global contextual modeling in integrating task-relevant representations. However, removing LBE (along with SI) leads to a less severe decrease than that observed with the removal of GBE, but it is still notable that macro precision decreased by 3.99\%, macro recall by 3.71\%, and macro F1-score by 3.48\%. This indicates that fundamental regional features shared across paradigms within each brain region also play a crucial role in enhancing the our model’s representation capability. Furthermore, removing only the SI module resulted in marginal performance reductions, less than 1\% across all three metrics, suggesting that although SI contributes to our model's generalisation ability, its impact is relatively limited compared to LBE and GBE.

These results demonstrate that the GBE and LBE modules effectively complement each other in capturing hierarchical representations at different levels. The LBE captures fundamental regional features shared across paradigms within each brain region, while the GBE focuses on integrating fundamental regional features into high-level feature representations associated with each task. Together, they significantly enhance our model’s generalization ability and overall performance.

\subsection{Parameter Sensitivity}
\begin{figure}[!t]
\centerline{\includegraphics[width=\columnwidth]{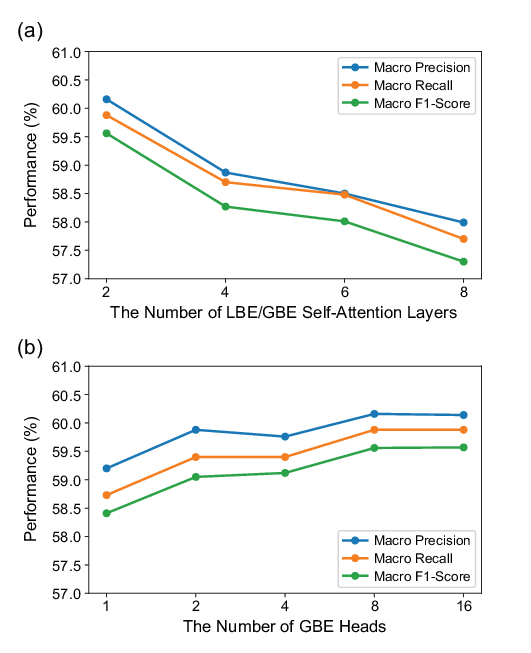}}
\caption{Parameter sensitivity analysis of the proposed Dual Local-Global Encoder (DLGE) model on classification performance. (a) Effect of varying the number of self-attention layers in both the local brain encoder (LBE) and global brain encoder (GBE). The model performs best with two layers, with performance gradually degrading as the number increases, likely due to overfitting. (b) Effect of changing the number of attention heads in GBE’s multi-head self-attention. Performance varies marginally, with the best results observed at eight heads, indicating that excessive heads offer limited benefit in this context.}
\label{fig:para}
\end{figure}

We explored the impact of important architectural parameters of the proposed DLGE model, which are the number of self-attention layers in both the LBE and GBE and the number of attention heads in GBE’s multi-head self-attention. In Transformer-based architectures, the depth of attention layers and the number of heads play a critical role in balancing representational capacity and generalization. To investigate the impact of architectural depth, we varied the number of layers in both LBE and GBE components across the range {2, 4, 6, 8}. Additionally, we examined the sensitivity of the multi-head attention mechanism by increasing the number of heads in GBE from 1 to 16, with results presented in Fig.~\ref{fig:para}.

As illustrated in Fig. 4(a), the increase in the number of attention layers leads to a decline in the classification performance. The best performance is achieved when the number of attention layers are set to two, which outperformed the worst case with 8 layers by approximately 2.17\%, 2.18\%, and 2.26\% in terms of macro precision, macro recall, and macro F1-score, respectively. This suggests that a shallow architecture is sufficient for capturing relevant dependencies in our study, while deeper stacks may lead to overfitting and reduced generalization.

In contrast, the performance change is relatively small  when the number of attention heads varied (see Fig. 4(b)). The best choice is eight heads, which surpasses the worst case of 1 head by approximately 0.96\%, 1.15\%, and 1.15\% in terms of macro precision, macro recall, and macro F1-score, respectively. This indicates that while multi-head attention contributes to marginal gain, the use of more heads beyond a fixed number offers limited additional benefit in this context.

\subsection{Visualization}

\begin{figure*}[htbp]
\centerline{\includegraphics[width=0.9\textwidth]
{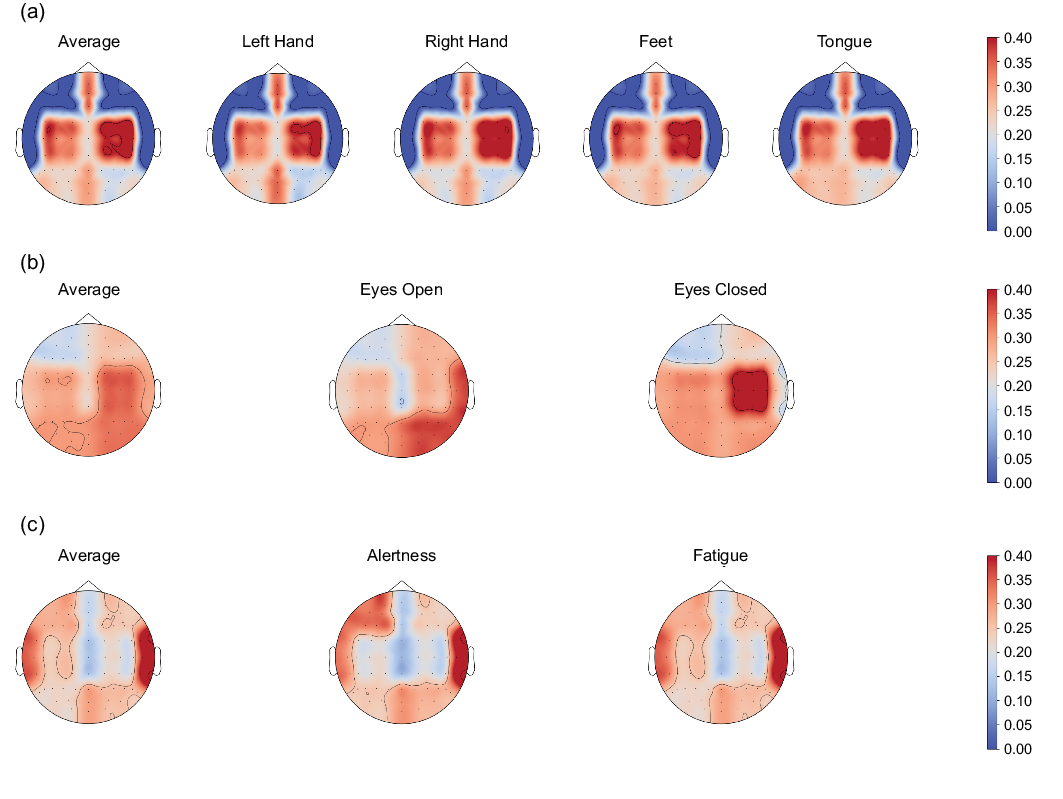}}
\caption{Topographical distribution of task-specific brain region contributions based on Gradient-weighted Class Activation Mapping (Grad-CAM). The leftmost column shows the average contribution map across all classes, while the remaining columns are class-specific contribution maps. (a) Motor Imagery Paradigm. Strong activation in the motor cortex across all classes. (b) Resting State Paradigm. “Eyes open” highlights right temporal and parietal-occipital regions, “eyes closed” emphasizes right motor cortex. (c) Driving Fatigue Paradigm. Right temporal region dominates and left frontal region is more active in the alertness state.}
\label{fig:vis}
\end{figure*}

To investigate how different brain regions contribute to classification performance across paradigms and classes, we applied Gradient-weighted Class Activation Mapping (Grad-CAM) \cite{selvaraju_grad-cam_nodate} for visualization. This method computes the gradients of the target class score with respect to the input embeddings or intermediate representations, generating class-discriminative localization maps that highlight the most influential brain regions. The resulting visualizations, shown in Fig.~\ref{fig:vis}, illustrate how the model attends to distinct brain regions for different paradigms and classes.

The motor imagery paradigm results in Fig. 5(a) show that the most significant contributions are concentrated in the motor cortex region, particularly in the right motor cortex region, which consistently shows higher importance than the left. The parietal-occipital region contributes significantly less, and all four classes in this paradigm exhibit a highly consistent importance distribution centered on the motor cortex region, in agreement with the shared policy-related areas across tasks.

For the resting state paradigm shown in Fig. 5(b), activity predominantly localizes to the right hemisphere, especially the right motor cortex. The left parietal region shows the lowest contribution across both classes. Specifically, in the “eyes open” class, the right temporal and right parietal-occipital regions show the highest contribution, whereas in the “eyes closed” class, the right motor cortex becomes the dominant region. This pattern may reflect a shift in internal cognitive processing which is with eyes closed, the lack of visual input may increase reliance on motor-related imagination, enhancing motor cortex activity and in contrast, with eyes open, external visual information remains available, reducing motor-related activation and increasing contributions from temporal and parietal-occipital regions associated with relaxation.

Fig. 5(c) presents the performance on the driving fatigue paradigm that the temporal regions, especially the right temporal region, show the strongest contributions. This is followed by the left frontal region, which shows higher importance in the alertness state. In the fatigue state, the left frontal contribution decreases notably.

Overall, the Grad-CAM visualization confirms that our model is able to focus on physiologically relevant brain regions under different tasks. For instance, motor imagery tasks yield high contributions from the motor cortex, while fatigue detection relies more heavily on parietal and occipital regions, aligning with established neuroscience findings \cite{pfurtscheller_motor_1997, chin-teng_lin_eeg-based_2005}. Resting-state conditions show a distinct shift in dominant regions depending on whether the eyes are open or closed.

Furthermore, the integration of dual-level encoder framework to learn more stable and task-related representations. The Grad-CAM visualization confirms that our model is able to focus on physiologically relevant brain regions under different tasks. For instance, motor imagery tasks yield high contributions from the motor cortex, while fatigue detection relies more heavily on parietal and occipital regions, aligning with established neuroscientific findings \cite{pfurtscheller_motor_1997, chin-teng_lin_eeg-based_2005}. Resting-state conditions show a distinct shift in dominant regions depending on whether the eyes are open or closed.

\section{Conclusions}

We conducted an initial study to investigate the feasibility of the cross-BCI-paradigm classification. A novel model DLGE is proposed to classify different paradigms without retraining. In this study, we also provide a solution to address two significant challenges (i.e., channel heterogeneity and interference between tasks) to enable cross-BCI-paradigm classification. The proposed anatomical inspired brain-region partitioning and padding strategy significantly enhances model compatibility across paradigms with varying channel configurations. The dual-level encoder architecture enables DLGE to learn stable high-level task-related representations. Through evaluation across three different paradigms, the results demonstrate that DLGE is feasible and effective for cross-BCI-paradigm classification. Ablation and sensitivity analyses confirm the contributions and robustness of each component, while Grad-CAM visualizations reveal task-relevant neural activation patterns. While effective, DLGE faces challenges in maintaining computational efficiency, suggesting future directions for lightweight and adaptive implementations.

\bibliographystyle{plain}
\bibliography{reference}
\end{document}